# Achievement of practical level critical current densities in PIT $Ba_{1-x}K_xFe_2As_2$/Ag tapes by conventional cold mechanical deformation

Zhaoshun Gao, Kazumasa Togano, Akiyoshi Matsumoto, and Hiroaki Kumakura*

National Institute for Materials Science, Tsukuba, Ibaraki 305-0047, Japan

**Abstract:**

We found that transport critical current density ($J_c$) of the $Ba_{1-x}K_xFe_2As_2$/Ag tapes is significantly enhanced by the combination process of cold flat rolling and uniaxial pressing. At 4.2 K, the $J_c$ reaches a high value of $9.1 \times 10^4$ A/cm$^2$ in 4 T, which is almost the practical level of $10^5$ A/cm$^2$. The $J_c$-$H$ curve shows extremely small magnetic field dependence and keeps a high value of $6.9 \times 10^4$ A/cm$^2$ in 10 T. These are the highest values reported so far for iron based superconducting wires. We also show our successful attempts of the fabrication of multi-filamentary tapes. A high $J_c$ value of $5.3 \times 10^4$ A/cm$^2$ (4.2 K, 10 T) was obtained in seven-filamentary tapes. The hardness and microstructure investigations reveal that the superior $J_c$ in uniaxial pressed tape samples is due to the high core density, more textured grains and the change of microcrack structure. These results indicate that iron based superconductors are promising for high magnetic field applications.

* Author to whom correspondence should be addressed; E-mail: KUMAKURA.Hiroaki@nims.go.jp

Since the discovery of superconductivity in LaFeAsO$_{1-x}$F$_x$, several types of iron based superconductors have been discovered [1-5]. Among them, K-doped (AE)Fe$_2$As$_2$ (AE = Sr, Ba. 122 type) is most potentially useful for high field applications due to their high critical current temperature ($T_c$) value of ~39 K, upper critical field ($H_{c2}$) of over 50 T and relatively small anisotropy [5-8]. Furthermore, the critical angle ($\theta_c$) of the transition from a strong link to a weak link for Ba122 is substantially larger than that of YBCO-based conductors [9]. Recently Co-doped Ba122 coated conductors have been grown by several groups utilizing the existing YBCO coated conductor technology and have reached a self-field $J_c$ over 1 MA/cm$^2$ [10-12]. Although at an early stage of development the transport $J_c$ in iron based superconductors reported was disappointingly low due to the weak link grain boundaries problem [13-26], astonishing progress has been made on Ba(Sr)122 wires in the past 3 years. The $J_c$ of Ba(Sr)122 wires can be easily approach the level 10$^4$ A/cm$^2$ at 4.2 K and 10 T through metal addition plus rolling induced texture process, hot isostatic press method, cold press method and so on [27-35]. Those results demonstrated that the mechanical deformation process is critical for producing high quality superconducting wires, which plays an important role of densifying the conductor core and aligning the grains of the superconducting phase. However, understanding how to further improve $J_c$ in Ba(Sr)122 by process optimization is still quite limited because we still lack a clear picture about the relationships between processing, microstructure, and superconducting properties. An understanding of the influence of mechanical deformation on the microstructure and superconducting properties will accelerate the development of appropriate process and further improve the transport $J_c$ of the Ba(Sr)122 wires. In this work, a comparative study was carried out between pressing and rolling mechanical processes. The variations in grain alignment, core density, microstructure and $J_c$ in the tapes were systematically investigated. We found that the uniaxial cold pressing is very useful to achieve the practical level $J_c$ in applied magnetic fields.

**Experiment**

The precursors of $Ba_{0.6}K_{0.4}Fe_2As_{2.1}$ were prepared from the Ba filings, K plates, Fe powder and As pieces. The details of fabrication process were described elsewhere [35]. After heat treatment, the precursor was then ground into powder using an agate mortar in a glove box filled with a high purity argon gas. The powder was packed into a Ag tube (outside diameter: 6 mm, inside diameter: 4 mm), which was subsequently groove rolled into a wire with the rectangular cross section of ~2 mm × ~2mm. The wires were deformed into a tape form using a flat rolling machine initially into 0.8 mm in thickness followed by the intermediate annealing at 800 ºC for 2 h and then into 0.40 ~ 0.20 mm in thickness. For pressed samples, the tape was then cut into 35 mm in length and uniaxially pressed between two hardened steel dies under the pressure of 2~3 G Pa. The rolled and pressed tapes were then subjected to the final heat treatment of 850 ºC for 5 h for sintering. All heat treatment was carried out by putting the samples into a stainless tube whose both ends were pressed and sealed by arc welding in an Ar atmosphere. We also fabricated seven-filamentary tapes using similar process. Mono-core wire with the diameter of ~1.3 mm was cut into seven pieces, bundled and put into another Ag tube. And then the assemblage was subjected to deformation, intermediate annealing, and final heat treatment by the same process as mono-core tape.

The transport current $I_c$ at 4.2 K and its magnetic field dependence were evaluated by standard four-probe method, with a criterion of 1 μV/cm. Magnetic fields up to 12 T were applied parallel to the tape surface. Vickers hardness was measured on the polished cross sections of the tape samples with 0.05 kg load and 10 s duration. We carried out mechanical polishing using emery paper and lapping paper, and then Ar ion polishing by cross section polisher of IB-09010CP (JEOL Co. Ltd.) to observe the surface morphologies of the tapes precisely. After the polishing, we performed scanning electron microscopy (SEM) observations using a SU-70 (Hitachi Co. Ltd.).

## Results & discussion

The phase impurity and texture of Ba122 tapes have been investigated by XRD analysis of the core surface, for which the silver sheath has been mechanically removed after cutting the tape sides. Figure 1 shows XRD patterns of rolled and pressed tapes. As a reference, the data of randomly orientated powder is also included in the figure. As can be seen, all samples consist of a main phase, $Ba_{1-x}K_xFe_2As_2$, but some Ag impurity phase coming from sheath material is also detected. The relative intensities of (*00l*) peaks with respect to that of (*103*) peak in all tape samples, when compared to random oriented powder, are strongly enhanced, indicating a relevant *c*-axis texture. However, the relative intensities of (*00l*) peaks almost remain same level in rolled tapes, suggesting that the grain texture is hardly further improved by rolling process in our samples. On the contrary, stronger relative intensity of (00*l*) peaks were observed in cold pressed tape. It indicates that cold pressing is more effective to improve the grain alignment. However, it should be noted that the degree of texture in our pressed tapes is still lower than that in Fe sheathed PIT Sr122 [29, 31]. It indicates that the texture could be further enhanced by optimization of processing parameters or using harder sheath materials.

Fig. 2 presents the field dependent transport $J_c$ of rolled and pressed tapes. The inset is the cross-section of the tape. The figure clearly shows that the transport $J_c$ significantly increases when reducing the rolling thickness. The $J_c$ achieved maximum value of $4.5 \times 10^4 A/cm^2$ at 10 T in the 0.26 mm thick tapes. When rolling the tape to smaller thickness, the degradation in critical current density was observed. However, it is worth noting that further improvement in $J_c$ values was achieved by the application of uniaxial pressing. All pressed tapes show a very weak field dependence as observed in the rolled tapes and the $J_c$ over $5.0 \times 10^4 A/cm^2$ at 10 T, indicating that a high $J_c$ is obtained with good reproducibility. The highest $J_c$ values obtained so far are $9.1 \times 10^4 A/cm^2$ at 4 T and $6.9 \times 10^4 A/cm^2$ at 10 T, respectively. Extrapolation of the $J_c$-H curve shows that the $J_c$ value at 2 T exceeds the practical level of $10^5 A/cm^2$. Even for the seven-filamentary tape, it still sustains a $J_c$ as high as $5.3 \times 10^4 A/cm^2$ at 10 T. Those $J_c$ values are the highest ever reported for the iron based superconducting wires so far, and highlight the importance of uniaxial pressing for enhancing the $J_c$ of iron based superconductors.

Because of difficulty of directly measuring the density of the thin superconducting core, people usually use Vickers hardness as an indication of the density of the core [36, 37]. In this work, we estimated the density of the core from Vickers hardness and investigated the relationships among the core density, the fabrication process, and $J_c$ values. Fig. 3a displays the influence of flat rolling on the hardness and $J_c$. This figure gives clear information about layer thickness dependence of $J_c$. The $J_c$ increases with decreasing tape thickness, reaches a maximum $4.5 \times 10^4$ A/cm$^2$ at 10 T for the 0.26 mm thick tapes. As illustrated in XRD, there is almost no difference in grain alignment for the rolled tapes. Therefore, the texture can be ruled out as a possible origin of the enhancement of transport $J_c$. However, an apparent difference in the hardness has been observed, the hardness increased with the progress of the rolling process. This indicates that the density might be the main reason of the $J_c$ enhancement in rolled tapes. The further rolling to smaller tape thicknesses caused a rapid reduction of hardness and $J_c$, which might be due to the microcracks or sausaging effect. As we know, when the tapes were rolled into very small thickness, the working instability became a serious problem [38]. In order to further improve the $J_c$, the technology of rolling to very thin tapes should be optimized in the future. Fig. 3b shows the $J_c$ (10T, 4.2 K) as a function of hardness for rolled and pressed tapes. It can be seen that there is a strong linear relation between the hardness of the tapes and $J_c$. As the hardness increased, the $J_c$ of the Ba122 core also increased, but the hardness and the $J_c$ of rolled tapes did not surpass the hardness and $J_c$ of the pressed tapes. This suggests that the uniaxially pressed samples perform much better than the rolled samples and also confirms our previous report that the high $J_c$ requires a high density Ba122 core [34, 35].

Figure 4 exhibits the typical SEM images of the polished surface for the rolled and pressed tapes. The observation was carried out on the tape plane of the tapes. It can be seen that although the rolling can reduce the voids and improve the density of Ba122 core, the microstructures are still porous and quite inhomogeneous. On the contrary, the pressed tapes with higher hardness and $J_c$ appeared to have a denser and uniform microstructure than the rolled tapes with lower hardness and $J_c$. This result is consistent with the hardness analysis. Generally, the force in pressing is larger and more uniform than in rolling. Thus, the microstructure in pressed sample is denser and more uniform than that in rolled one.

Among the many stages of fabrication, flat rolling has been commonly used to densify and align the superconducting core [29, 31, 39]. Our results showed a large increase of $J_c$ due to the improvement of the core density and preferred orientation in the initial step by rolling process. For further rolling to smaller tape thicknesses, the degradation in critical current density was observed. On the contrary, when the tape was pressed, the $J_c$ were significantly increased by further improving the core density and grain alignment. Many authors have already demonstrated the advantage of pressing in Bi-based tapes [40, 41]. This advantage can be attributed to the change of crack structure [42] and more uniform deformation achieved during pressing than rolling. During the rolling, the pressure varies along the arc of contact between the rolls and the sample. The stress induced by inhomogeneous pressure will promote cracks to align transversely to the length direction, thus blocking the current flow. However, in the case of uniaxial pressing, the stress situation is rotated 90° with respect to the tape normal resulting in cracks along the direction of the tape length as shown in Fig. 4d. In addition, the forces applied by uniaxial pressing are perpendicularly and uniformly distributed on the surface of the sample, thus resulting in higher and homogeneous compression. The higher and uniform pressure reduces the voids, improves texture formation, and thus further improves the $J_c$. However, it should be emphasized that practical application of uniaxial pressing process to manufacture of long length wires require specialized machine for continuous pressing the tape. Fortunately, there have been successful attempts at production of long Ag/Bi-2223 wires by periodic pressing [43] and eccentric rolling [44], which might be also applied for the production of long length Ba122 wires with high transport $J_c$.

## Conclusion

Excellent transport $J_c$ values of $\sim 10^5 \text{A/cm}^2$ under magnetic fields were obtained in uniaxial pressed $Ba_{1-x}K_xFe_2As_2$ tapes. A comparative study of microstructure and hardness between pressing and rolling mechanical processes indicates that the high core density, more aligned grains and the change of microcrack structure are responsible for this high $J_c$ performance. With further improvement of critical current density and wire fabrication technology, use of $Ba_{1-x}K_xFe_2As_2$ for high field application is feasible.

*Note added:* During preparation of our manuscript, a recent report also demonstrated that high $J_c$ can be obtained in Sr122/Ag tapes by hot pressing [45]. This result supports our conclusion that the high core density is crucial for high $J_c$ performance in iron based superconductor.

## Acknowledgements

This work was supported by the Japan Society for the Promotion of Science (JSPS) through its "Funding Program for World-Leading Innovative R&D on Science and Technology (FIRST) Program. We acknowledge Dr. H. Fujii, Mr. S. J. Ye and Miss Y. C. Zhang of the National Institute for Materials Science for their assistance in $I_c$ measurement.

# Captions

Figure 1   X-ray diffraction patterns for $Ba_{1-x}K_xFe_2As_2$ random powder, rolled and pressed tapes fabricated in different deformation process.

Figure 2   The transport $J_c$ values obtained in this experiment plotted as a function of applied magnetic fields.

Figure 3   (a) the influence of flat rolling on the hardness and $J_c$. (b) the $J_c$ (10 T, 4.2 K) as a function of hardness for rolled and pressed tapes.

Figure 4   The SEM surface images of flat-rolled tape with thickness of 0.39mm (a) and 0.26 mm (b), and pressed tape (c). The crack structure for pressed tape (d).

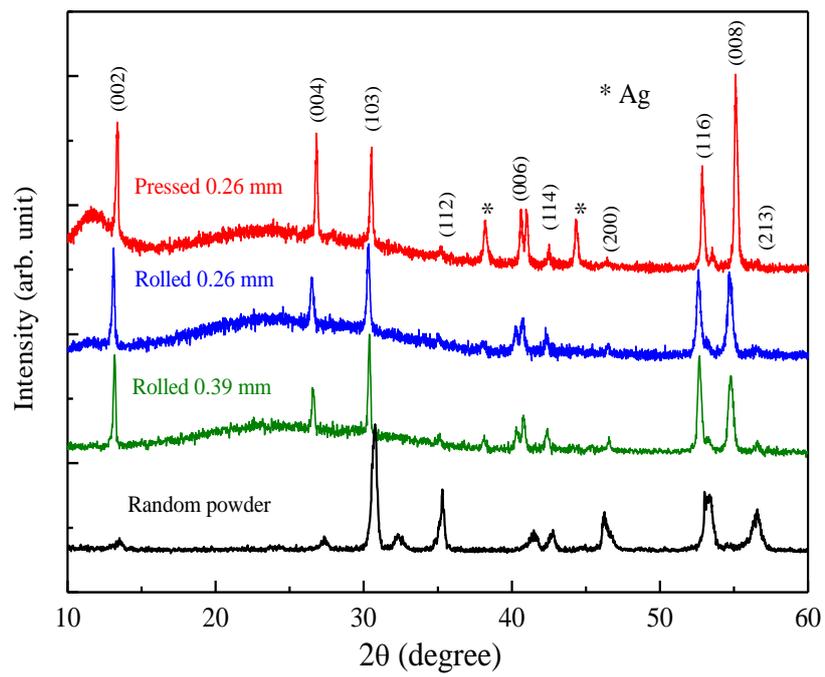

Fig.1 Gao et al.

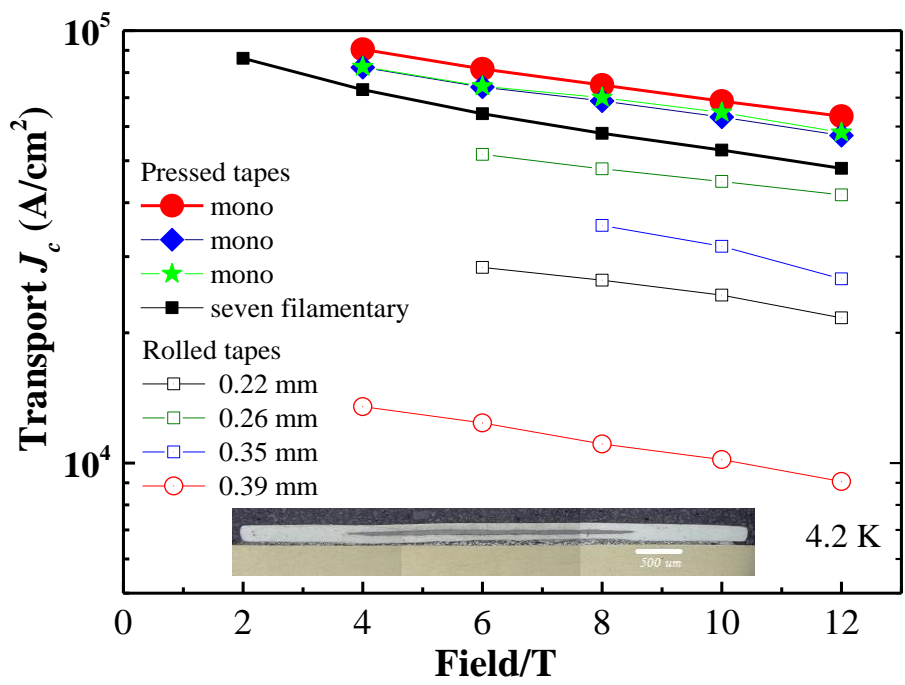

Fig.2 Gao et al.

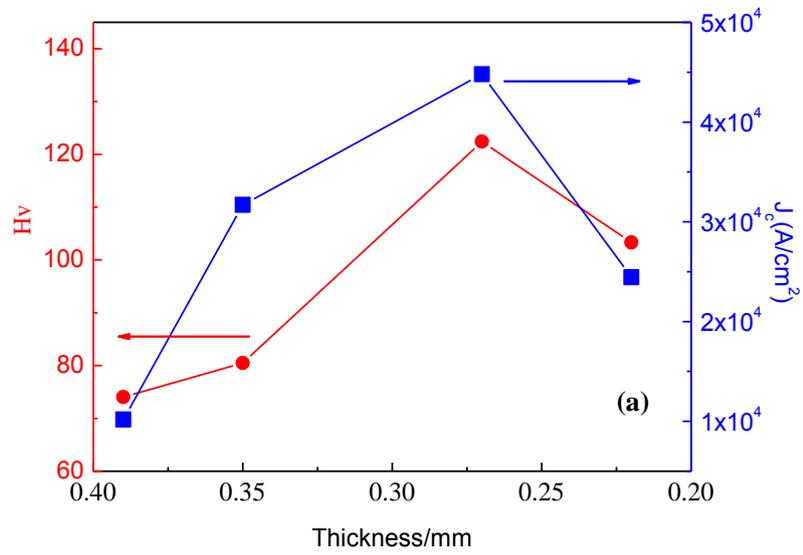

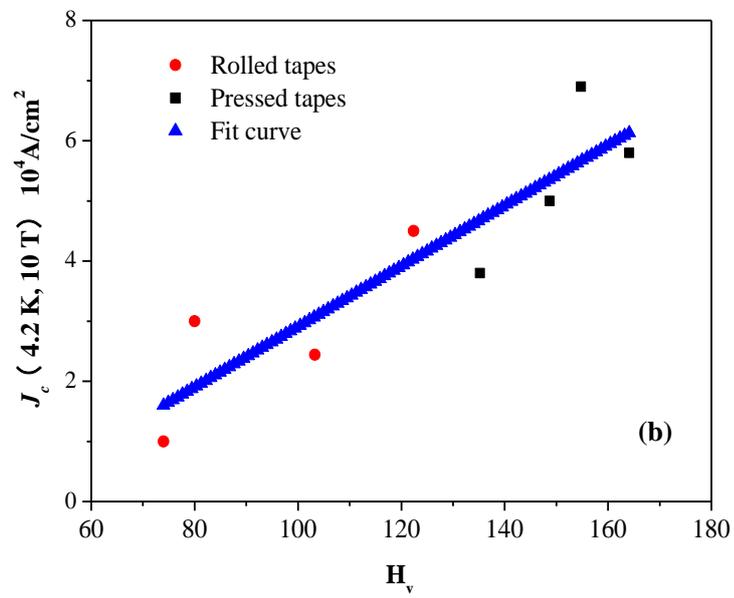

Fig.3 Gao et al.

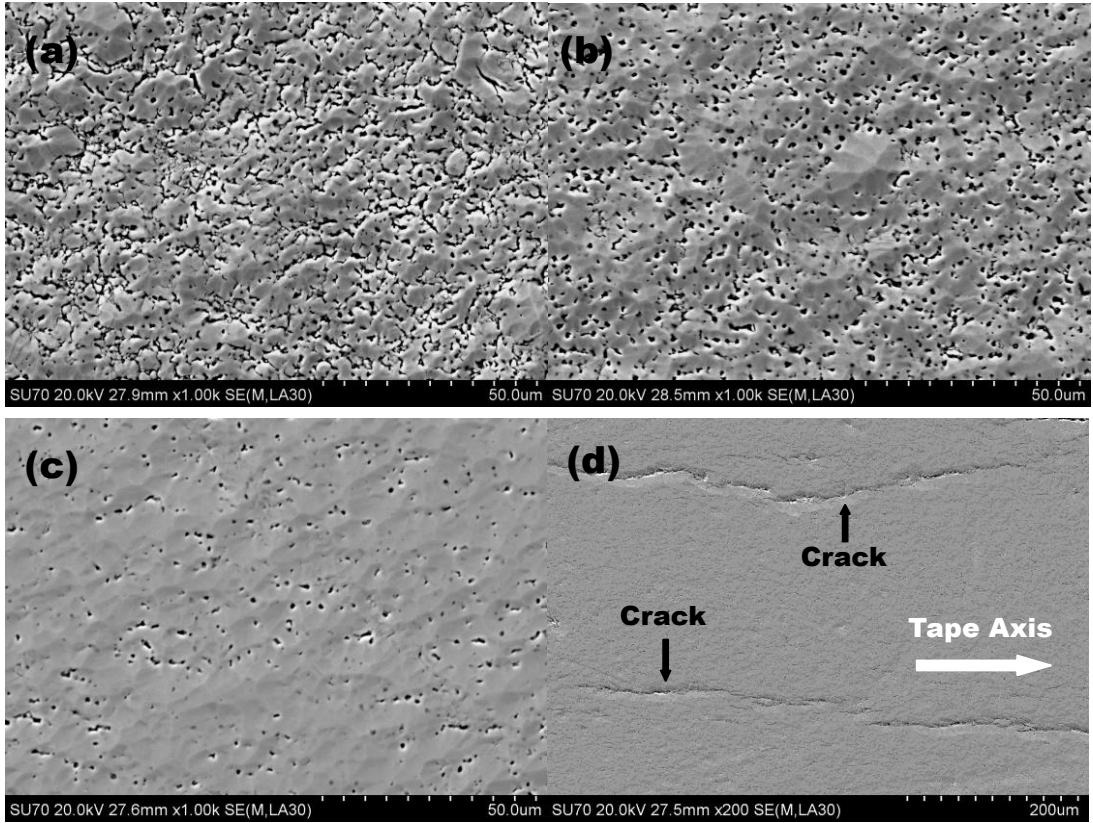

Fig.4 Gao et al.